\documentclass[preprint,showpacs,preprintnumbers,amsmath,amssymb]{revtex4}

\usepackage{graphicx}% Include figure files
\usepackage{dcolumn}% Align table columns on decimal point
\usepackage{bm}% bold math

\begin{document}
\preprint{APS/123-QED}

\title{Surface nanobubbles as a function of gas type}% Force line breaks with \\
\author{Michiel A. J. van Limbeek$^1$}
\author{James R. T. Seddon$^1$}
\email{j.r.t.seddon@utwente.nl}
\affiliation{$^1$Physics of Fluids, MESA+ Institute for Nanotechnology, University of Twente, P.O. Box 217, 7500 AE Enschede, The Netherlands}

\begin{abstract}
We experimentally investigate the nucleation of surface nanobubbles on PFDTS-coated silicon as a function of the specific gas dissolved in the water.  In each case we restrict ourselves to  equilibrium conditions ($c=100\,\%$, $T_{liquid} = T_{substrate}$).  Not only is nanobubble nucleation  a strong function of gas type, but there also exists an optimal system temperature of $\sim 35-40\,\mathrm{^oC}$ where nucleation is maximized, which is weakly dependent on gas type.  We also find that contact angle is a function of nanobubble radius of curvature for all gas types investigated.  Fitting this data allows us to describe a line tension which is dependent on the type of gas, indicating that the nanobubbles are sat on top of adsorbed gas molecules.  The average line tension was $\tau  \sim -0.8 \mathrm{nN}$.
\end{abstract}
%\pacs{}
\maketitle

\section{Introduction}
Surface nanobubbles are nanoscopic gaseous domains that exist on the solid/liquid interface~\cite{parker1994,lou2000,tyrrell2001,holmberg2003,steitz2003,simonsen2004,zhang2004,borkent2007,yang2008,brenner2008,seddon2011}.  They are surprisingly stable ~\cite{zhang2008,brenner2008,ducker2009,seddon2010b}, surviving orders of magnitude longer than the classical diffusive life time.  Since their recent discovery, several key observations have been made regarding the `ideal' conditions for nucleation~\cite{zhang2006d,yang2007,seddon2010c}.  These include the (almost~\cite{hampton2010,comment_liquid}) uniqueness for the formation in water, or in solutions containing at least some part water.  On the other hand, changing the substrate chemistry can lead to differences in the typical nanobubble sizes and distributions~\cite{simonsen2004,agrawal2005,kameda2008,borkent2009}, but nanobubbles are always found as long as sufficient care is used as well as an appropriate experimental technique/pretreatment~\cite{lou2000,tyrrell2001,lou2002,holmberg2003,simonsen2004,switkes2004,zhang2006,kameda2008a}.

One area that has seemingly escaped detailed research is the dependence of nanobubble formation on the \textit{type} of gas that is dissolved in the water.  Almost all studies to date investigate nanobubbles made from `air', with the few exceptions including carbon dioxide~\cite{yang2003,yang2007,zhang2008}, butane~\cite{miller1999,kameda2008a}, and nitrogen~\cite{kameda2008a} (we omit here the hydrogen and oxygen nanobubbles created using electrolysis~\cite{zhang2006b,yang2009,zhang2010}, since  gas was not necessarily dissolved in the water).    Of  these specific gases, carbon dioxide and butane seem peculiar choices.  Both of these gases have triple points above room temperature so can liquefy if the pressure is sufficiently high, i.e. if they fill a nanobubble which is sufficiently small.  As an example, carbon dioxide was used in Reference~\cite{zhang2008} because of its strong rotational fine structure signal, which allowed use of the attenuated total reflection infrared spectroscopy technique to show conclusively that nanobubbles contain gas.  However the critical pressure and temperature of CO$_2$ are $p_c=7.4\,\mathrm{MPa}$ and $T_c=31.0\,\mathrm{^oC}$~\cite{crc}, respectively, so we would expect liquefaction to occur in these CO$_2$-filled nanobubbles if their radii of curvature was lower than $R_c = 2\gamma/(p_c-p_0)\approx 20\,\mathrm{nm}$ (note that the nanobubbles in Reference~\cite{zhang2008} were much larger than this).  In butane-saturated water,  \textit{nanodroplets} have already been reported on Si(100) by Kameda \textit{et al.}~\cite{kameda2008}.

It is clear that gas type plays an important role, but no study has ever been carried out to investigate the effects of different gases.   Thus,  we have directly investigated the role of gas type on the formation of surface nanobubbles.  We chose a selection of different gases, including noble gases (helium and argon), diatomic gases (hydrogen, nitrogen, and oxygen), and more complex gases (carbon dioxide and methane).  Furthermore, for each gas type we restricted ourselves to systems where the liquid was saturated  with the gas and specifically \textit{not} supersaturated, following recent observations that supersaturation is not a requirement for nanobubble formation~\cite{seddon2010c}.  Thus our observations provide information on surface nanobubble nucleation in  equilibrium.

\section{Experimental method}
The substrate was a silicon wafer that had been hydrophobised with a self assembled monolayer of perfluorodecyltrichlorosilane (PFDTS), following the guidelines of Reference~\cite{helmutthesis,seddon2010a}.  In brief, we degassed the chamber that contained the uncoated substrate to a
pressure of $\sim50\,\mathrm{\mu bar}$ (i.e. below the vapor pressure of the PFDTS).  Next, we opened this chamber to a reservoir of degassed PFDTS for $5\,\mathrm{min}$
so that the silane molecules adsorbed onto the substrate.  Finally, we closed
the system off to the PFDTS reservoir and opened it to a reservoir of degassed
pure water for half an hour, hence increasing the pressure of the system
to the vapor pressure of water to allow the reagents to react.
The resulting rms roughness was $0.4\,\mathrm{nm}$, and the contact angles were $\sim 110\,\mathrm{^o}$ (equilibrium), $\sim 116\,\mathrm{^o}$ (advancing), and $\sim 104\,\mathrm{^o}$ (receding).  This surface has no known phase transitions in the temperature range investigated in the present study.

The substrate was mounted on a temperature controlled sample plate (331 temperature controller, Lakeshore, USA) before a purpose-built atomic force microscope (AFM) liquid cell was firmly mounted on top.  The AFM was an Agilent 5100 operated in tapping mode with a typical scan
speed of $4-5\,\,\mathrm{\mu m/s}$.  The AFM cantilevers were hydrophilic, Au-back-coated Si$_3$N$_4$ Veeco NPG probes, with typical
spring constants and resonance frequencies in water of $0.58\,\mathrm{N/m}$ and $25-35\,\mathrm{kHz}$.  We operated the AFM at a set point of $90\,\%$.

For the liquid we used ultrapure water (Simplicity 185 purification system, Millipore SAS, France).  This was placed within a stainless steel container that included a pump-out port, a re-pressurization port, and a digital pressure gauge.

Our study was to investigate the role of the gas type  on nanobubble nucleation, so we now explain the procedure that we used  to prepare the gas dissolved in the water.  The first stage of preparation was to thoroughly degas the water to remove as much air as possible.  During this stage the pressure was reduced to $\lesssim 20\,\mathrm{mbar}$ and, once this pressure was reached,  we continued to pump for at least a further $\sim 30\,\mathrm{mins}$.  The container of water was then closed off to the pump and placed within a heat bath which was at the correct temperature for the experiment at hand.

\begin{figure}
\begin{center}
\includegraphics[width=8cm]{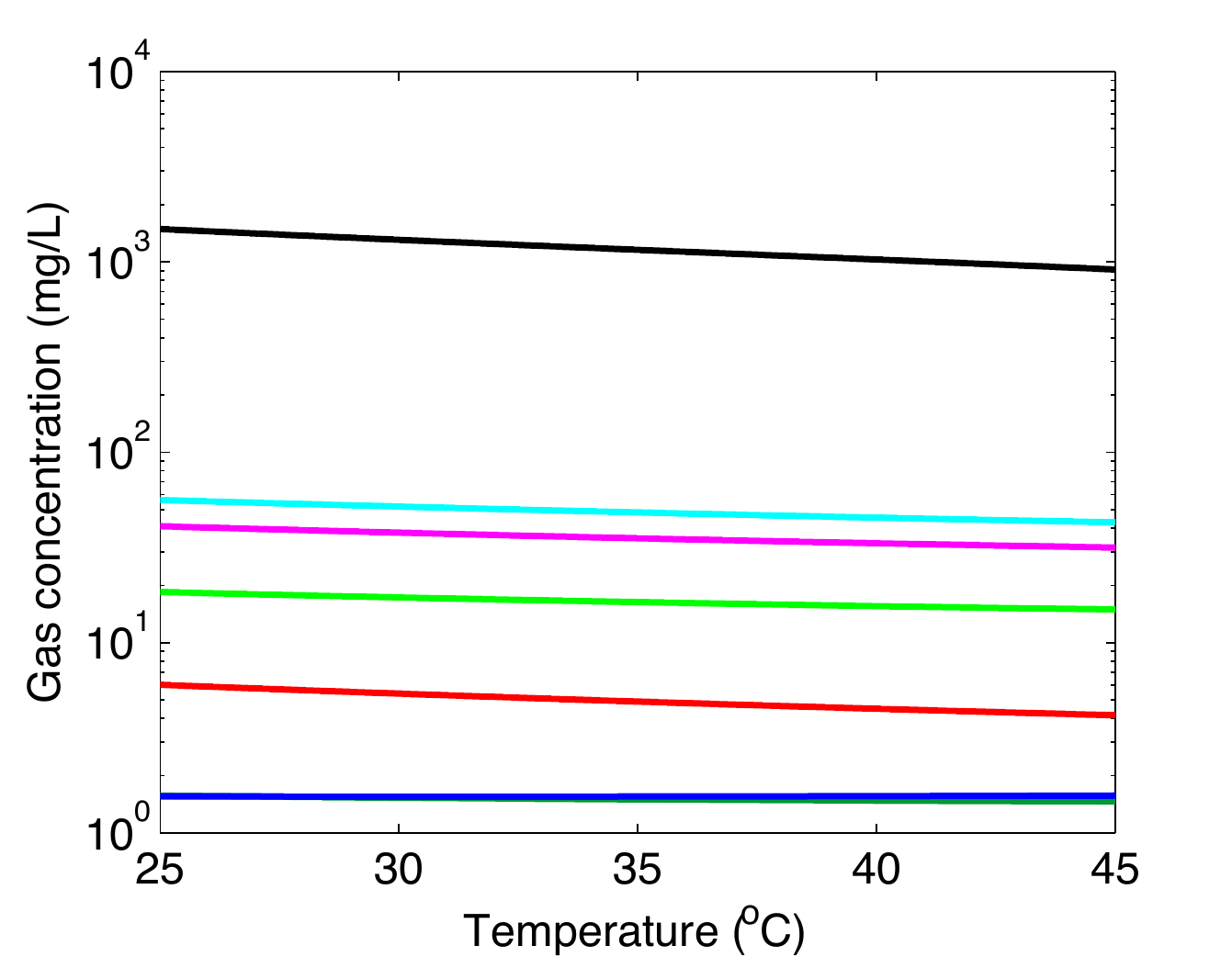}
\caption{Concentration of saturated gases in water as a function of temperature.  Colors are for H$_2$ (dark green), He (blue), CH$_4$ (red), N$_2$ (green), O$_2$ (magenta), Ar (cyan), and CO$_2$ (black).  Note that the data for H$_2$ and He overlie each other. \label{fig:mgperl}}
\end{center}
\end{figure}

The stainless steel container had a volume of $100\,\mathrm{mL}$ but we only treated $40\,\mathrm{mL}$ of water at a time.  This allowed us to repressurise the evacuated volume above the water level with a specific type of gas.  The number of molecules of a specific type that dissolve in water at equilibrium is well known\cite{crc},  thus the magnitude of the  repressurisation was to $1\,\mathrm{atm}$ plus the required number of molecules for dissolution multiplied by $kT/V_w$, where $kT$ is the thermal energy and $V_w$ is the $40\,\mathrm{mL}$ volume of water.  We plot the solubilities of the different gases used in this study, as functions of temperature, in ~\ref{fig:mgperl}.  After the container had been pressurized we closed it off to the gas supply and  maintained the system's seal for $2-3$ gas-diffusion time scales.

Once the liquid was prepared it was injected into the liquid cell of our atomic force microscope using a syringe pump.  The syringe was at the same temperature as both the water and the substrate.  The AFM was placed inside a glass environment-control chamber that had had its air displaced  with one atmosphere of the same gas type that was dissolved in the liquid.  Thus, any gas exchange between the liquid and the environment would \textit{not} lead to cross contamination of the gas type.  This means that the gas content of the water was maintained throughout  the measurements and our results provide information of nanobubble nucleation in  fully equilibrated systems.

\section{Results and discussion}
\begin{figure}
\begin{center}
\includegraphics[width=15cm]{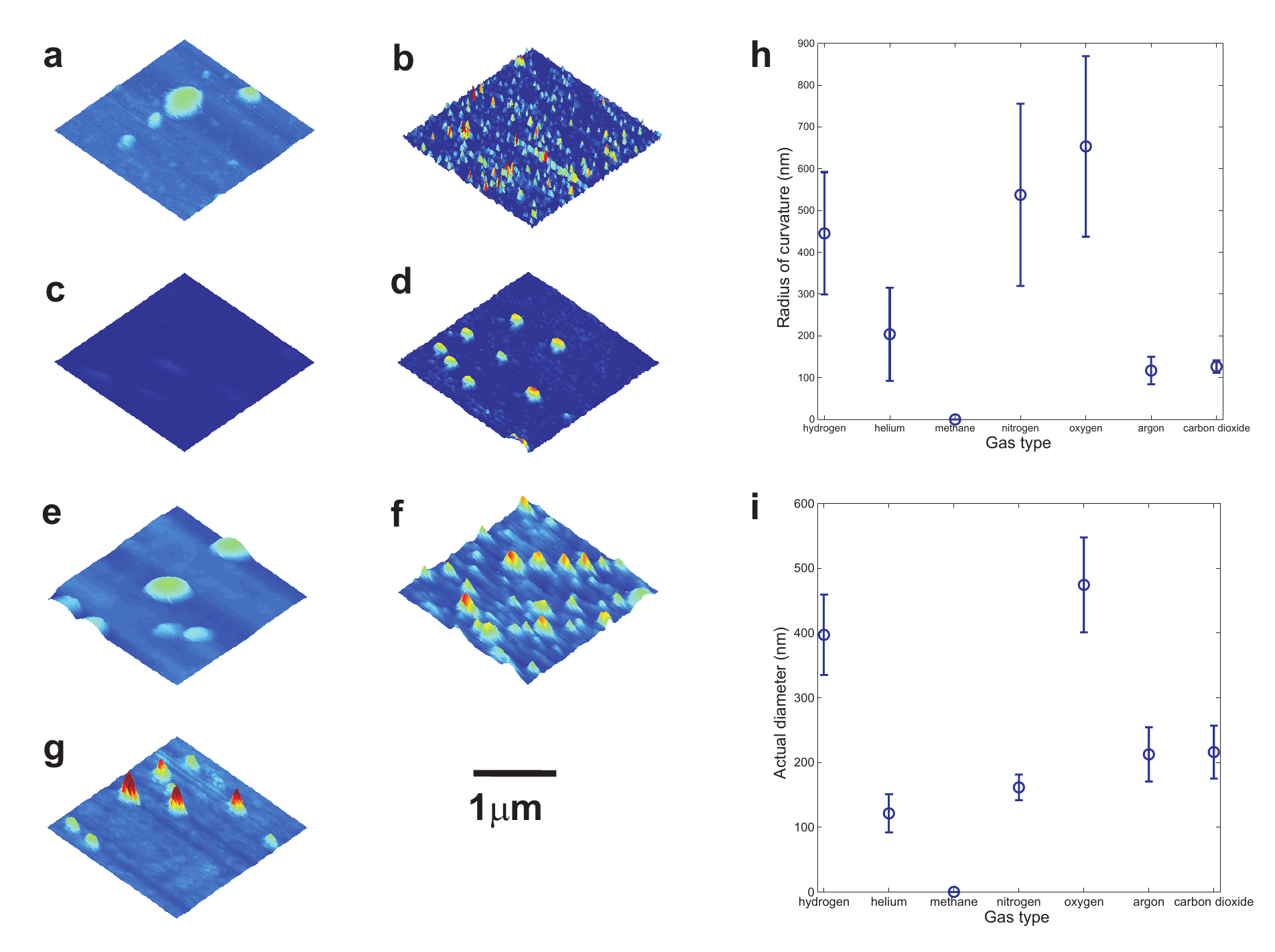}
\end{center}
\caption{Typical images of nanobubbles formed from different gas types at $25\,\mathrm{^oC}$. Gases are (a) H$_2$, (b) He, (c) CH$_4$, (d) N$_2$, (e) O$_2$, (f) Ar, and (g) CO$_2$.  Images are  $2\,\mathrm{\mu m} \times 2\,\mathrm{\mu m}$; height scales are (b,d) $20\,\mathrm{nm}$ and (a,c,e-g) $250\,\mathrm{nm}$. (h) Average radii of curvature and (i) average diameters of nanobubbles versus gas type (error bars correspond to one standard deviation). \label{fig:25deg}}
\end{figure}

We present typical images of nanobubbles nucleated from different gases at $25\,\mathrm{^oC}$ in  ~\ref{fig:25deg}.  Each image represents $2\,\mathrm{\mu m} \times 2\,\mathrm{\mu m}$, but the  height scales differ as described in the caption.  Each gas type resulted in distinct nanobubble sizes/densities, with methane being the only gas that led to zero nucleation at this temperature.  Note that each set of experimental conditions was repeated  in their entirety, including the water preparation stage, and  for a given experiment several areas of the substrate were scanned.

\begin{figure}
\begin{center}
\includegraphics[width=8cm]{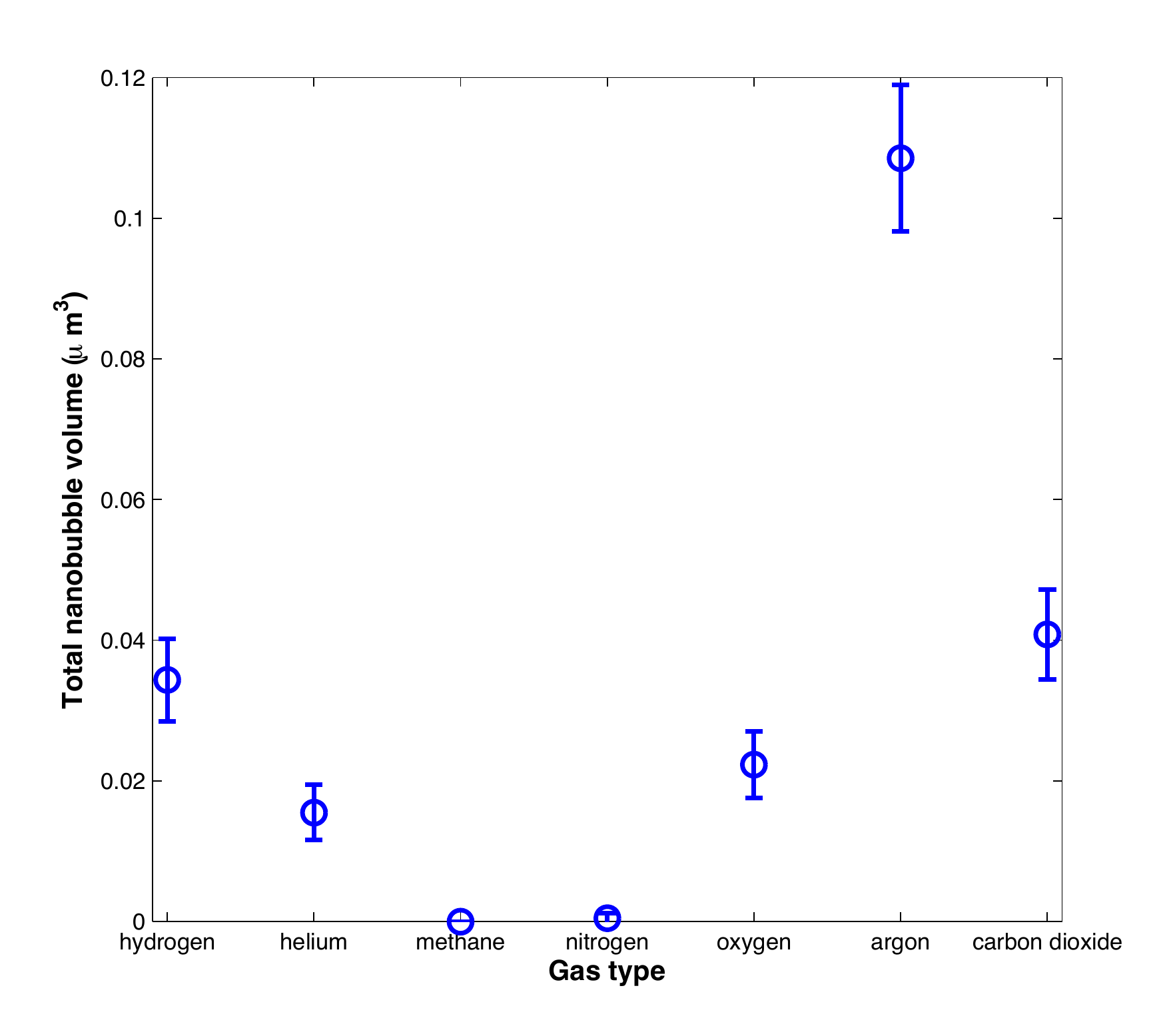}
\end{center}
\caption{The total nanobubble volume on a $2\,\mathrm{\mu m} \times 2\,\mathrm{\mu m}$ area as a function of gas type at a system temperature of $25\,\mathrm{^oC}$.  The abscissa is ordered with increasing molecular weight and solubility.  If gas solubility was the control parameter governing nucleation we would expect the data in this graph to vary monotonically. \label{fig:density25}}
\end{figure}

To quantify the differences between the gas types we plot the average radii of curvature and diameters of nanobubbles as  functions of gas type in ~\ref{fig:25deg}h and ~\ref{fig:25deg}i, respectively,  as well as the calculated total volume of nanobubbles on a $2\,\mathrm{\mu m} \times 2\,\mathrm{\mu m}$ area in  ~\ref{fig:density25}, where we plot the `gas type' along the abscissa in order of increasing molecular weight and solubility   (note that each data point is an \textit{average} of several different areas and several experimental runs - we are  displaying the average value as a `standard' measure of nanobubble populations, see References \cite{zhang2005,seddon2010c}).  If solubility was the  control parameter governing nanobubble nucleation, we would expect the data in  ~\ref{fig:density25} to be a monotonically varying function.  Clearly this is not the case, with no obvious functional form able to describe the data.  This means that solubility is \textit{not} the governing factor for nanobubble nucleation in a system that is  in equilibrium (i.e. with the substrate and liquid temperatures being equal, and the gas saturated within the liquid).

Secondly, although adsorption strengths of the various gases are not known for our substrate, we  expect the ordering of these to be He and H$_2$ (weak); Ar, O$_2$, and N$_2$ (medium); CH$_4$ and CO$_2$ (strong) \cite{dunne1996,staudt2005,sudik2005}.  Clearly there is no direct dependence on this either.  Hence we can  rule out the possibility of nanobubbles \textit{solely} nucleating from the bulk desorption of dense adsorbates, which may exist in the form of micropancakes \cite{seddon2010a}.

Another possible origin of the nucleation may be the excess gas that swells the density-depleted layer immediately next to an immersed hydrophobic solid \cite{doshi2005,dammer2006}.  However, the dependency of the amount of swelling on gas type is currently disputed \cite{doshi2005,mezger2006,sendner2009}.  Thus none of the three most likely possible causes for nanobubble nucleation, namely formation from gas dissolved in the bulk, formation from molecules adsorbed on the substrate, or formation from the gas-enriched layer near the hydrophobic substrate, can be individually responsible.

In order to try to understand the gas dependency of nanobubble nucleation further, we proceeded to investigate the effects of different temperatures on nucleation.  For this, we repeated the experiments for methane, nitrogen, and oxygen, this time at system temperatures of $30\,\mathrm{^oC}$, $35\,\mathrm{^oC}$, $40\,\mathrm{^oC}$, and $45\,\mathrm{^oC}$.

\begin{figure*}
\begin{center}
\includegraphics[width=16cm]{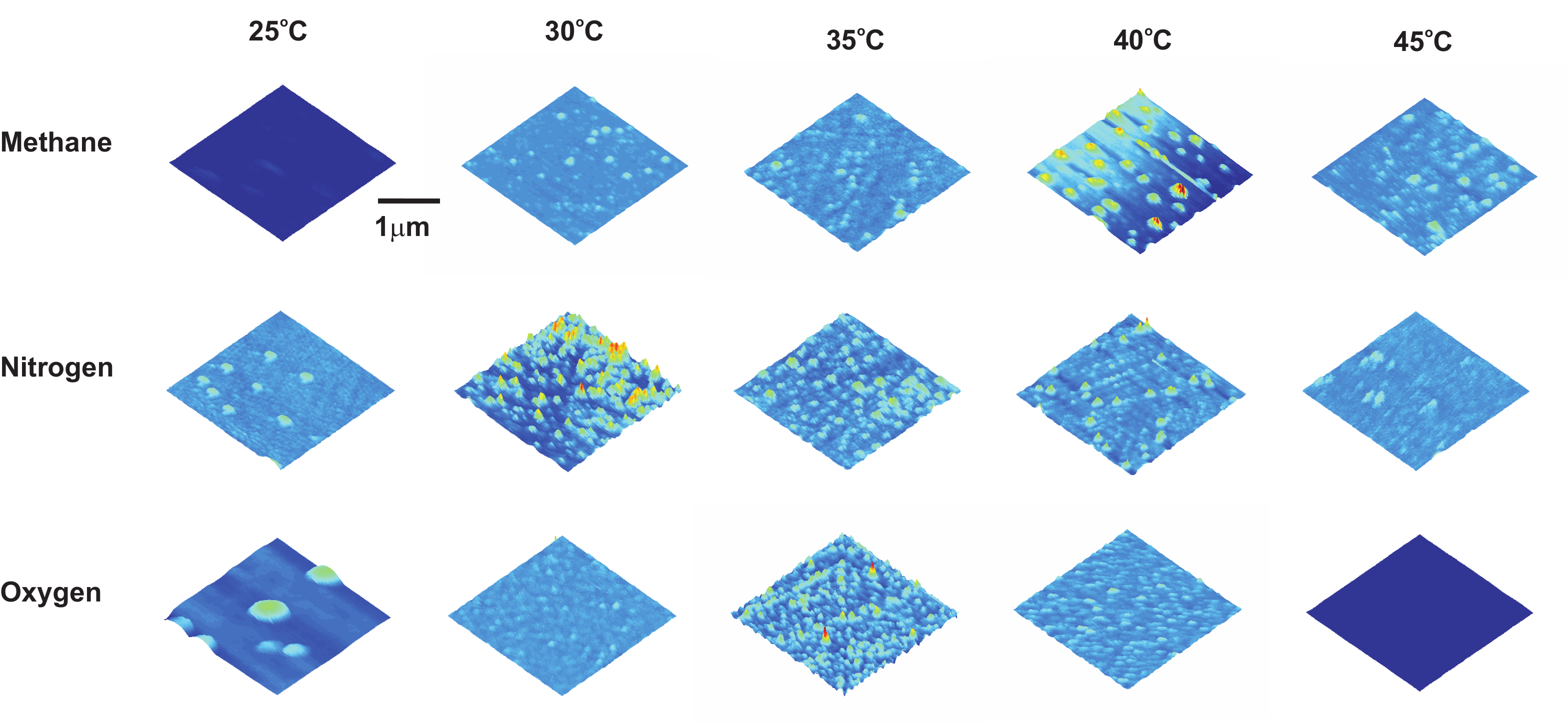}
\end{center}
\caption{Temperature dependence of nanobubbles formed from methane, nitrogen, and oxygen-saturated water at temperatures of (a) $25\,\mathrm{^oC}$, (b) $30\,\mathrm{^oC}$, (c) $35\,\mathrm{^oC}$, (d) $40\,\mathrm{^oC}$, and (e) $45\,\mathrm{^oC}$.  Images are $2\,\mathrm{\mu m} \times 2\,\mathrm{\mu m}$.  Height scales are (methane, $40\,\mathrm{^oC}$) $230\,\mathrm{nm}$, (methane, other) $40\,\mathrm{nm}$; (nitrogen) $40\,\mathrm{nm}$; (oxygen, $25\,\mathrm{^oC}$) $250\,\mathrm{nm}$, (oxygen, other) $40\,\mathrm{nm}$.\label{fig:ramp}}
\end{figure*}

\begin{figure}
\begin{center}
\includegraphics[width=8cm]{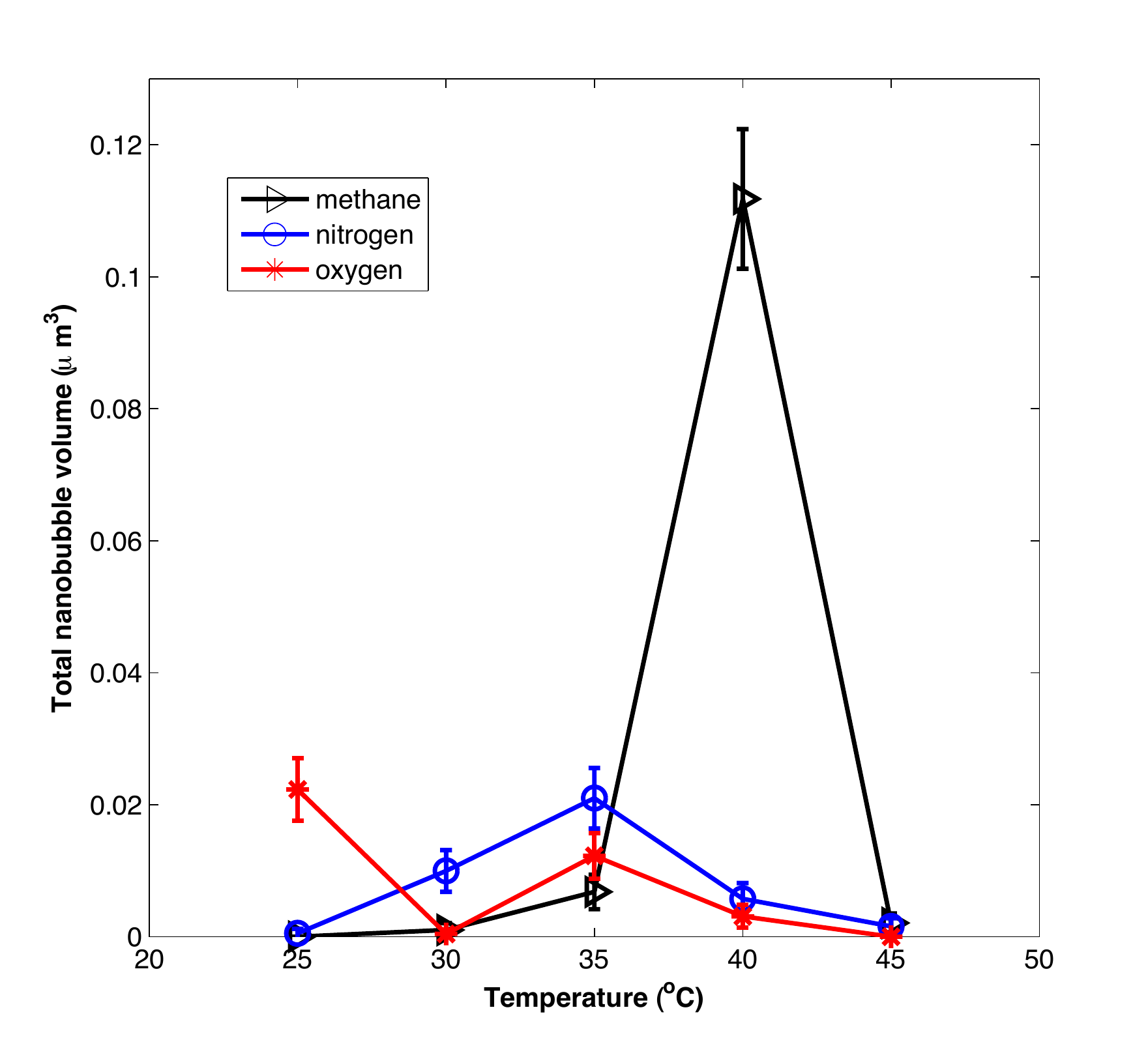}
\end{center}
\caption{The total nanobubble volume on a $2\,\mathrm{\mu m} \times 2\,\mathrm{\mu m}$ area as a function of gas type and temperature for  methane (triangles), nitrogen (circles), and oxygen (asterisks).  There is a clear maximum in density between $35$ and $40\,\mathrm{^oC}$ for all three gas types.\label{fig:volumes}}
\end{figure}

Typical images of nanobubbles created from the three gases, as  functions of system temperature, are shown in  ~\ref{fig:ramp}.  It is evident that an optimal temperature exists for each type of gas that leads to maximum nanobubble nucleation.  As an example, no nanobubbles nucleated from the methane-saturated water at $25\,\mathrm{^oC}$, but nucleation occurred for every temperature we tested above this, with maximum production at $40\,\mathrm{^oC}$.  These optimal temperatures are more clearly visible in the plots of the total nanobubble volume in  ~\ref{fig:volumes}.  Other work~\cite{zhang2005,seddon2010c} has described a similar maximum in \textit{air}-filled nanobubbles  as being due to a maximum in solubility.  However, if we compare the curves of  ~\ref{fig:volumes} to the temperature dependencies of the solubilities  in  ~\ref{fig:mgperl}, it is clear that there is no corresponding peak in solubility, i.e. we reiterate that solubility is \textit{not} the control parameter for nanobubble nucleation in gas-equilibrated systems.

\begin{figure}
\begin{center}
\includegraphics[width=8cm]{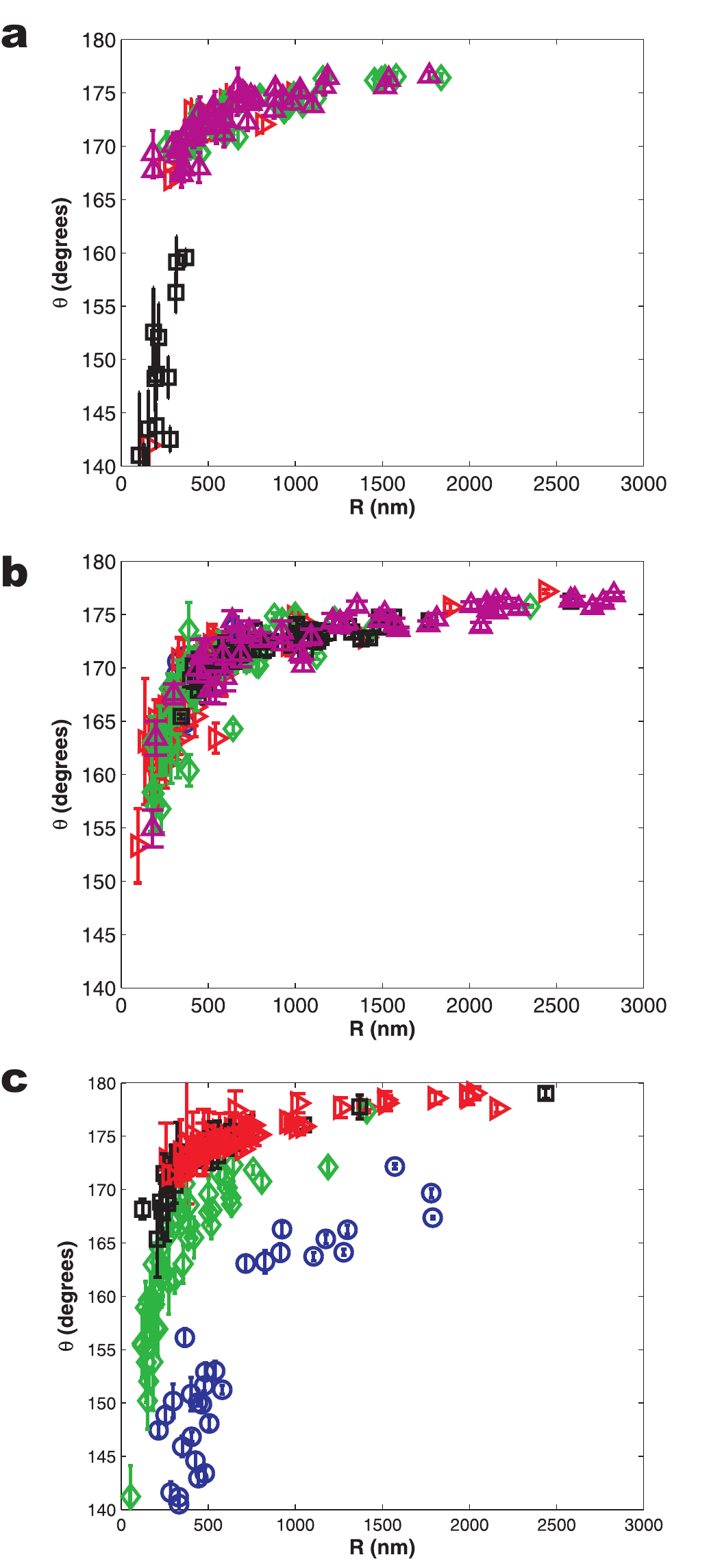}
\caption{Contact angle as a function of nanobubble radius of curvature for (a) methane, (b) nitrogen, and (c) oxygen nanobubbles at $25\,\mathrm{^oC}$ (circles), $30\,\mathrm{^oC}$ (right triangles), $35\,\mathrm{^oC}$ (diamonds), $40\,\mathrm{^oC}$ (squares), and $45\,\mathrm{^oC}$ (up triangles).  Dependence of contact angle on radius of curvature indicates that line tension must be considered.  Data has been corrected for cantilever tip distortion. \label{fig:th_R_3}}
\end{center}
\end{figure}

In total, over 300 nanobubbles were imaged for the present study, so we have sufficient statistics to extract information for the contact angle versus radius of curvature.  We plot this for methane, nitrogen, and oxygen in  ~\ref{fig:th_R_3}, where we use different symbols and colors to represent the differing system temperatures.  For all of the gases in the present study the contact angle was found to increase with increasing radius of curvature, before leveling off to $\theta \rightarrow \sim 180\,\mathrm{^o}$ as $R \rightarrow \infty$.

\begin{figure}
\begin{center}
\includegraphics[width=8cm]{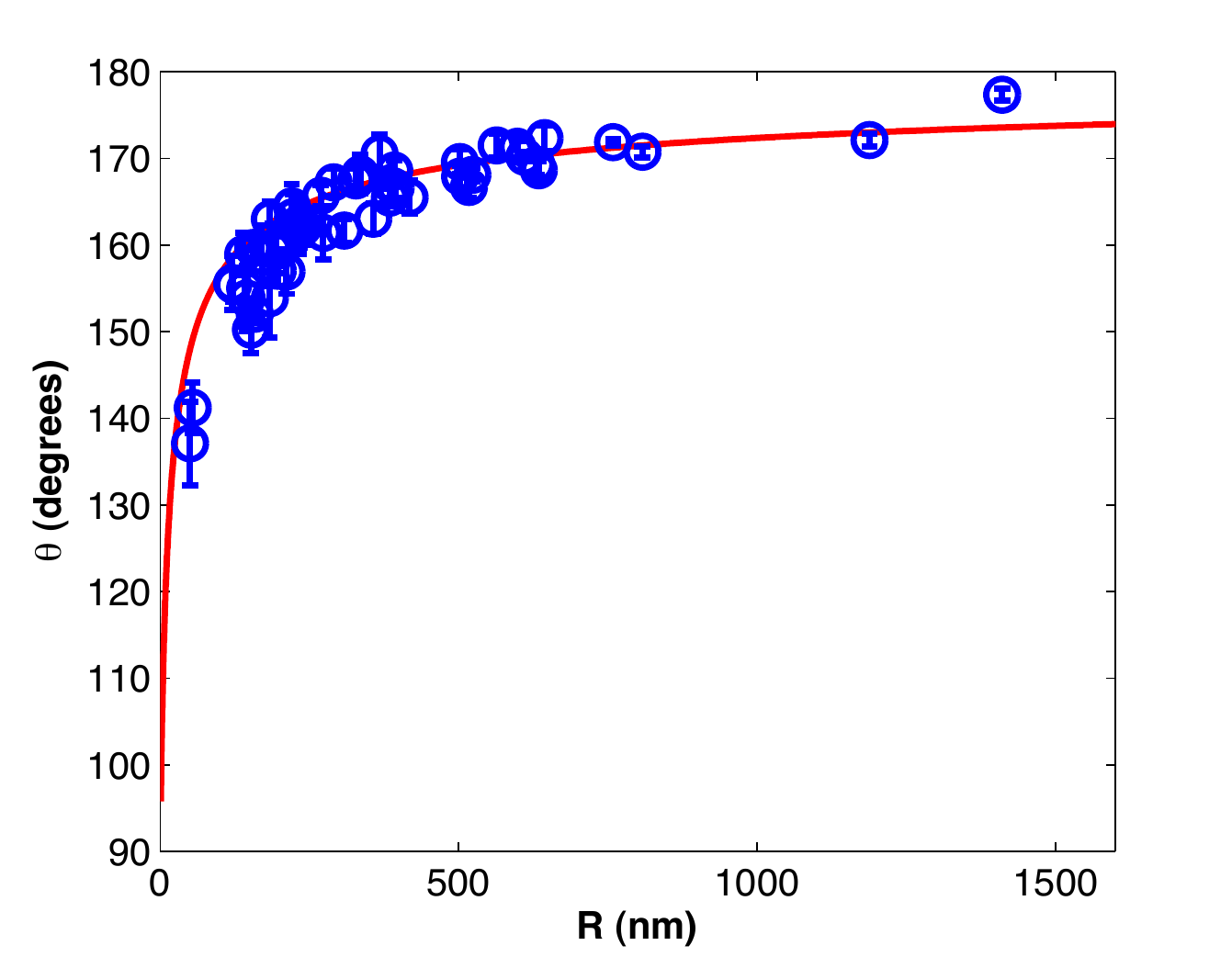}
\end{center}
\caption{Fit of the $35\,\mathrm{^oC}$ data for contact angle versus radius of curvature for oxygen.  Fit is of the form $\cos\theta = \cos\theta_\infty - (\cos\theta_\infty - \cos\theta_0)/(1+R/\delta)$, with $\theta_\infty=0\,\mathrm{^o}$, $\theta_0=90\,\mathrm{^o}$, and $\delta=9\,\mathrm{nm}$.  The corresponding line tension is $\tau = -\gamma\delta = -0.6\,\mathrm{nN}$.}
\label{oxygenfit}
\end{figure}

When contact angle is a function of the radius of curvature, it is necessary to introduce a line tension.  Brenner and Lohse~\cite{brenner2008} suggest a functional form for this effect of
\begin{equation}
\cos\theta=\cos\theta_\infty-\frac{\cos\theta_\infty-\cos\theta_0}{1+R/\delta}.
\end{equation}
Here, $\theta_\infty$ is the value of the contact angle at large radii, $\theta_0$ is at diminishingly small radii, and $\delta$ is the length scale over which we expect line tension to take effect.    Hence, we now choose to use the functional form of Reference~\cite{brenner2008} for the nanoscale correction to contact angle demonstrated in ~\ref{fig:th_R_3}.  For the three fitting parameters, $\theta_\infty$, $\theta_0$, and $\delta$, we set the two angles to $180\,\mathrm{^o}$ and $90\,\mathrm{^o}$, respectively, and use $\delta$ as the single fitting parameter for the data.   These choices for the angles were selected because (i) it is clear from the data in  ~\ref{fig:th_R_3} that the contact angle tends to $180\,\mathrm{^o}$ with increasing radius of curvature, and (ii) the contact angle rapidly reduces for smaller bubbles but we do not want or expect the nature of the material to alter from hydrophobic to hydrophilic with decreasing radius of curvature. Within this framework line tension would be $\tau = -\gamma \delta$, where $\gamma$ is the surface tension of water.  A typical fit is shown in \ref{oxygenfit}, where we present the  $35\,\mathrm{^oC}$ data for oxygen.  In this case, a line tension of $\tau \approx -0.6\,\mathrm{nN}$ becomes important at a length scale of $\delta \approx 9\,\mathrm{nm}$.

\begin{figure}
\begin{center}
\includegraphics[width=8cm]{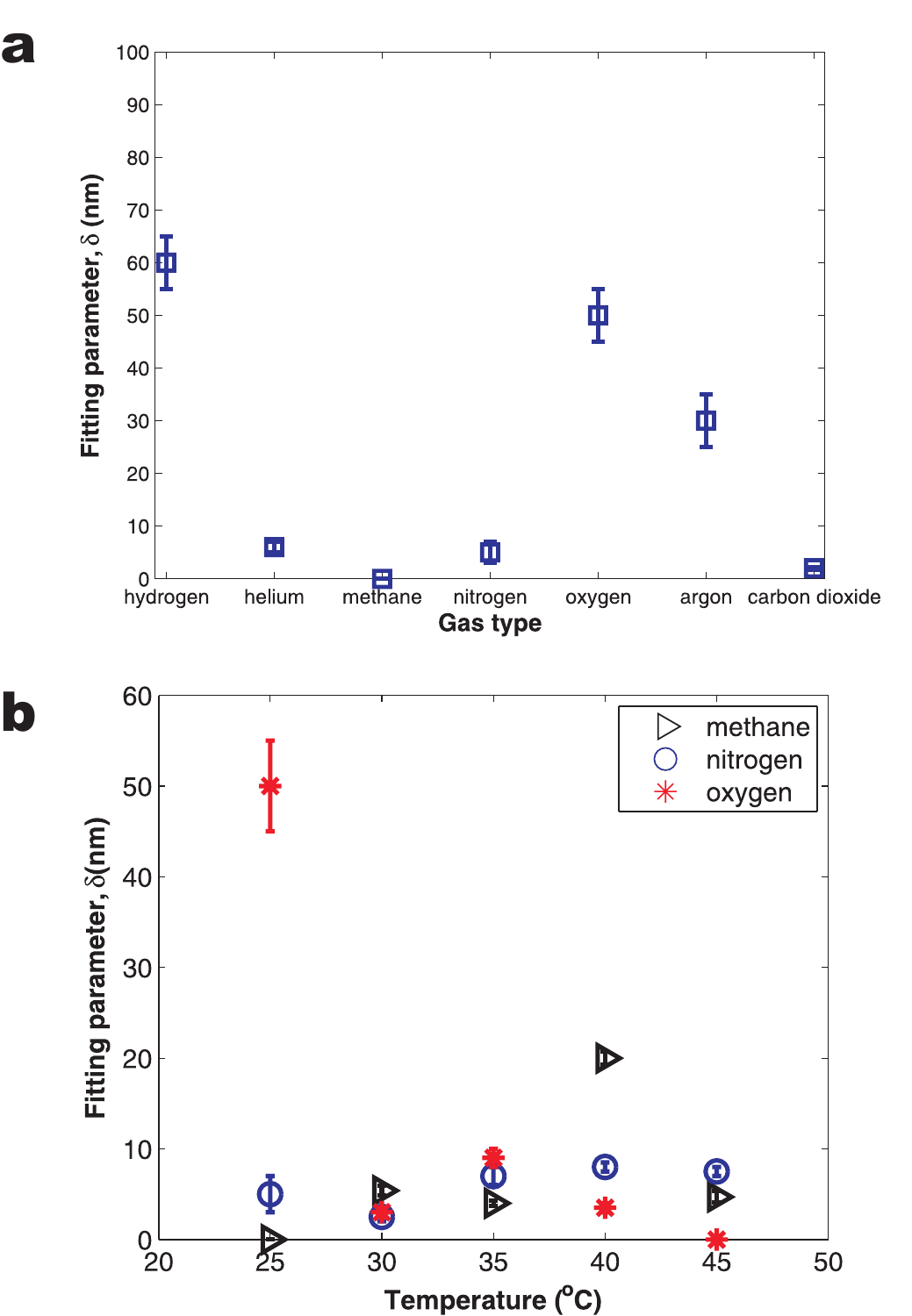}
\end{center}
\caption{Fitting parameter $\delta$ as (a) a function of gas type at a temperature of $25\,\mathrm{^oC}$ and (b) as a function of temperature for methane, nitrogen, and oxygen.  The average line tension is $\tau = -\gamma\delta \sim -0.8\,\mathrm{nN}$.}
\label{delta}
\end{figure}

Values of $\delta$ for each of the different gases used in this study, as well as for the different temperatures used for methane, nitrogen, and oxygen, are presented in ~\ref{delta}.  There is a large spread in $\delta$ with gas type, which gives insight into the possible configuration of the nanobubble-substrate geometry.  Different values of $\delta$ for different gas types means that the contact angle is different for different gas types.  This is puzzling since contact angle is dependent on the three differing surface energies of the three different interfaces (solid, liquid, and gas), but it is always the \textit{densest} phase that contributes the most to these values. (As an example, degassing the air above the water level in a glass beaker does not lead to a change in contact angle of the meniscus - the air has been evacuated, but the water and glass remain.)  Thus, the specific type of gas should not have a noticeable effect on contact angle because this is dominated by the water and solid.  The way to resolve this issue of variable contact angle with gas type would be if the nanobubble was sat on top of a dense adsorbate (``micropancake,'' \cite{zhang2007}) of gas molecules.  As pointed out by Reference ~\cite{seddon2011}, this is the most probable configuration for a nanobubble -- if a nanobubble is sat on top of a dense adsorbate, it is the binding energy of gaseous molecules to adsorbed molecules that should be considered and \textit{not} the binding energy to the underlying solid.  Not only does this explain the essentially constant value of contact angle for air-filled nanobubbles, regardless of the substrate, but it also explains the difference of contact angle with gas type.  Of course, an underlying adsorbate would also provide a gas bank for the gaseous influx to balance the diffusive outflux in the dynamic equilibrium model of nanobubble stability \cite{brenner2008}.

The mean value of line tension here, which we calculate as the average of the data in ~\ref{delta}a,b, is $\tau \sim -0.8\,\mathrm{nN}$.  Measurement of a negative line tension for nanobubbles/nanodroplets is expected\cite{weijs2010}.  Values in the literature include $-3\,\mathrm{nN}$ \cite{yang2003}, $\sim -2.3\,\mathrm{pN}$ \cite{checco2003}, $-0.2\,\mathrm{nN}$ \cite{kameda2008}, and $\approx -1\,\mathrm{pN}$ \cite{weijs2010}.

Returning to \ref{delta}, we can again reiterate that we find no clear functional form linking solubility or adsorption strength to the different values of $\delta$.  Thus, we posit the following possible route to nucleation, using the noble gases. Argon should adsorb more strongly to the substrate than helium, resulting in less argon nanobubbles, but the opposite is found. The fact that there is approximately an order of magnitude more argon available in the bulk, however, should increase the density of the adsorbate, possibly to include multiple layers. Thus nanobubble nucleation may occur as a result of bulk desorption from the more weakly bound \textit{upper} layers.  For the diatomic molecules (hydrogen, nitrogen, and oxygen) it is a lot less clear.  The strength of these adsorbates is dependent on the mean orientation of the molecules, and we would expect bulk desorption to occur much more readily than for the noble gases.  The same is also true for the more complex molecules (carbon dioxide and methane).

\section{Conclusions}
We have shown that gas type is a key parameter for the nucleation of nanobubbles.  Not only do specific gases lead to the formation of more nanobubbles than others, but there also exists an optimal temperature for nanobubble nucleation between $\sim 35\,\mathrm{^oC}$ and $\sim 40\,\mathrm{^oC}$, which appears to be weakly dependent on gas type.

Surprisingly, nanobubble nucleation is not directly dependent on either the solubility of the specific gas in water, or on the relative adsorption strength of the gas to the substrate.  This indicates that nanobubbles do not form solely due to the amount of gas available in the bulk, or from dense adsorbates (micropancakes) on the substrate.  Furthermore, we should expect either zero or very small dependence  on the specific gas type for the gas-enrichment layer thickness near a hydrophobic substrate, so this does not provide an adequate solution to nucleation either.

Hence, nanobubble nucleation must come from a combination of several competing factors.   Certainly (at least) a two-stage nucleation process is most likely.  On the one hand, more nanobubbles are found with increasing temperature for low temperatures, as would be consistent for systems reliant on desorption, whilst less nanobubbles are found with increasing temperature for high temperatures, now consistent with systems reliant on solubility.  The cross over in behaviour is at a temperature of $\sim 35-40\,\mathrm{^oC}$.

For all the nanobubbles thatwe investigated we found a dependence of contact angle on the radius of curvature.  We  introduced a line-tension term of the  form proposed by Brenner and Lohse~\cite{brenner2008}.  On our PFDTS-coated silicon substrate, the average line tension was negative and equal to $\tau  \sim -0.8 \mathrm{nN}$.

The authors acknowledge useful discussions with Detlef Lohse and Harold Zandvliet throughout the work.  The research leading to these results has received funding from the European Community's Seventh Framework Programme (\textit{FP7/2007-2013}) under \textit{grant agreement} number 235873, and from the Foundation for Fundamental Research on Matter (FOM), which is sponsored by the Netherlands Organization for Scientific Research (NWO).

%\bibliography{../../../nano_litrev_bib}

\begin{mcitethebibliography}{46}
\providecommand*\natexlab[1]{#1}
\providecommand*\mciteSetBstSublistMode[1]{}
\providecommand*\mciteSetBstMaxWidthForm[2]{}
\providecommand*\mciteBstWouldAddEndPuncttrue
  {\def\EndOfBibitem{\unskip.}}
\providecommand*\mciteBstWouldAddEndPunctfalse
  {\let\EndOfBibitem\relax}
\providecommand*\mciteSetBstMidEndSepPunct[3]{}
\providecommand*\mciteSetBstSublistLabelBeginEnd[3]{}
\providecommand*\EndOfBibitem{}
\mciteSetBstSublistMode{f}
\mciteSetBstMaxWidthForm{subitem}{(\alph{mcitesubitemcount})}
\mciteSetBstSublistLabelBeginEnd
  {\mcitemaxwidthsubitemform\space}
  {\relax}
  {\relax}

\bibitem[Parker et~al.(1994)Parker, Claesson, and Attard]{parker1994}
Parker,~J.~L.; Claesson,~P.~M.; Attard,~P. \emph{J. Phys. Chem.} \textbf{1994},
  \emph{98}, 8468--8480\relax
\mciteBstWouldAddEndPuncttrue
\mciteSetBstMidEndSepPunct{\mcitedefaultmidpunct}
{\mcitedefaultendpunct}{\mcitedefaultseppunct}\relax
\EndOfBibitem
\bibitem[Lou et~al.(2000)Lou, Ouyang, Zhang, Li, Hu, Li, and Yang]{lou2000}
Lou,~S.-T.; Ouyang,~Z.-Q.; Zhang,~Y.; Li,~X.-J.; Hu,~J.; Li,~M.-Q.; Yang,~F.-J.
  \emph{J. Vac. Sci. Technol. B} \textbf{2000}, \emph{18}, 2573--2575\relax
\mciteBstWouldAddEndPuncttrue
\mciteSetBstMidEndSepPunct{\mcitedefaultmidpunct}
{\mcitedefaultendpunct}{\mcitedefaultseppunct}\relax
\EndOfBibitem
\bibitem[Tyrrell and Attard(2001)Tyrrell, and Attard]{tyrrell2001}
Tyrrell,~J. W.~G.; Attard,~P. \emph{Phys. Rev. Lett.} \textbf{2001}, \emph{87},
  176104\relax
\mciteBstWouldAddEndPuncttrue
\mciteSetBstMidEndSepPunct{\mcitedefaultmidpunct}
{\mcitedefaultendpunct}{\mcitedefaultseppunct}\relax
\EndOfBibitem
\bibitem[Holmberg et~al.(2003)Holmberg, K\"uhle, Garnaes, Morch, and
  a.~Boisen]{holmberg2003}
Holmberg,~M.; K\"uhle,~A.; Garnaes,~J.; Morch,~K.~A.; a.~Boisen,
  \emph{Langmuir} \textbf{2003}, \emph{19}, 10510--10513\relax
\mciteBstWouldAddEndPuncttrue
\mciteSetBstMidEndSepPunct{\mcitedefaultmidpunct}
{\mcitedefaultendpunct}{\mcitedefaultseppunct}\relax
\EndOfBibitem
\bibitem[Steitz et~al.(2003)Steitz, Gutberlet, Hauss, Kl\"osgen, Krastev,
  Schemmel, Simonsen, and Findenegg]{steitz2003}
Steitz,~R.; Gutberlet,~T.; Hauss,~T.; Kl\"osgen,~B.; Krastev,~R.; Schemmel,~S.;
  Simonsen,~A.~C.; Findenegg,~G.~H. \emph{Langmuir} \textbf{2003}, \emph{19},
  2409--2418\relax
\mciteBstWouldAddEndPuncttrue
\mciteSetBstMidEndSepPunct{\mcitedefaultmidpunct}
{\mcitedefaultendpunct}{\mcitedefaultseppunct}\relax
\EndOfBibitem
\bibitem[Simonsen et~al.(2004)Simonsen, Hansen, and Kl\"osgen]{simonsen2004}
Simonsen,~A.~C.; Hansen,~P.~L.; Kl\"osgen,~B. \emph{J. Colloid Interface Sci.}
  \textbf{2004}, \emph{273}, 291--299\relax
\mciteBstWouldAddEndPuncttrue
\mciteSetBstMidEndSepPunct{\mcitedefaultmidpunct}
{\mcitedefaultendpunct}{\mcitedefaultseppunct}\relax
\EndOfBibitem
\bibitem[Zhang et~al.(2004)Zhang, Zhang, Lou, Zhang, Sun, and Hu]{zhang2004}
Zhang,~X.~H.; Zhang,~X.~D.; Lou,~S.~T.; Zhang,~Z.~X.; Sun,~J.~L.; Hu,~J.
  \emph{Langmuir} \textbf{2004}, \emph{20}, 3813--3815\relax
\mciteBstWouldAddEndPuncttrue
\mciteSetBstMidEndSepPunct{\mcitedefaultmidpunct}
{\mcitedefaultendpunct}{\mcitedefaultseppunct}\relax
\EndOfBibitem
\bibitem[Borkent et~al.(2007)Borkent, Dammer, Sch\"onherr, Vancso, and
  Lohse]{borkent2007}
Borkent,~B.~M.; Dammer,~S.~M.; Sch\"onherr,~H.; Vancso,~G.~J.; Lohse,~D.
  \emph{Phys. Rev. Lett.} \textbf{2007}, \emph{98}, 204502\relax
\mciteBstWouldAddEndPuncttrue
\mciteSetBstMidEndSepPunct{\mcitedefaultmidpunct}
{\mcitedefaultendpunct}{\mcitedefaultseppunct}\relax
\EndOfBibitem
\bibitem[Yang et~al.(2008)Yang, Kooij, Poelsema, Lohse, and
  Zandvliet]{yang2008}
Yang,~S.; Kooij,~E.~S.; Poelsema,~B.; Lohse,~D.; Zandvliet,~H. J.~W. \emph{EPL}
  \textbf{2008}, \emph{81}, 64006\relax
\mciteBstWouldAddEndPuncttrue
\mciteSetBstMidEndSepPunct{\mcitedefaultmidpunct}
{\mcitedefaultendpunct}{\mcitedefaultseppunct}\relax
\EndOfBibitem
\bibitem[Brenner and Lohse(2008)Brenner, and Lohse]{brenner2008}
Brenner,~M.~P.; Lohse,~D. \emph{Phys. Rev. Lett.} \textbf{2008}, \emph{101},
  214505\relax
\mciteBstWouldAddEndPuncttrue
\mciteSetBstMidEndSepPunct{\mcitedefaultmidpunct}
{\mcitedefaultendpunct}{\mcitedefaultseppunct}\relax
\EndOfBibitem
\bibitem[Seddon and Lohse(2011)Seddon, and Lohse]{seddon2011}
Seddon,~J. R.~T.; Lohse,~D. \emph{J. Phys. Cond. Mat.} \textbf{2011},
  \emph{23}, 133001\relax
\mciteBstWouldAddEndPuncttrue
\mciteSetBstMidEndSepPunct{\mcitedefaultmidpunct}
{\mcitedefaultendpunct}{\mcitedefaultseppunct}\relax
\EndOfBibitem
\bibitem[Zhang et~al.(2008)Zhang, Quinn, and Ducker]{zhang2008}
Zhang,~X.~H.; Quinn,~A.; Ducker,~W.~A. \emph{Langmuir} \textbf{2008},
  \emph{24}, 4756--4764\relax
\mciteBstWouldAddEndPuncttrue
\mciteSetBstMidEndSepPunct{\mcitedefaultmidpunct}
{\mcitedefaultendpunct}{\mcitedefaultseppunct}\relax
\EndOfBibitem
\bibitem[Ducker(2009)]{ducker2009}
Ducker,~W.~A. \emph{Langmuir} \textbf{2009}, \emph{25}, 8907--8910\relax
\mciteBstWouldAddEndPuncttrue
\mciteSetBstMidEndSepPunct{\mcitedefaultmidpunct}
{\mcitedefaultendpunct}{\mcitedefaultseppunct}\relax
\EndOfBibitem
\bibitem[Seddon and Zandvliet(2010)Seddon, and Zandvliet]{seddon2010b}
Seddon,~J. R.~T.; Zandvliet,~H. J.~W. \emph{Surf. Sci.} \textbf{2010},
  \emph{604}, 476--477\relax
\mciteBstWouldAddEndPuncttrue
\mciteSetBstMidEndSepPunct{\mcitedefaultmidpunct}
{\mcitedefaultendpunct}{\mcitedefaultseppunct}\relax
\EndOfBibitem
\bibitem[Zhang et~al.(2007)Zhang, Zhang, Sun, Zhang, Li, Fang, Xiao, Zeng, and
  Hu]{zhang2006d}
Zhang,~X.~H.; Zhang,~X.; Sun,~J.; Zhang,~Z.; Li,~G.; Fang,~H.; Xiao,~X.;
  Zeng,~X.; Hu,~J. \emph{Langmuir} \textbf{2007}, \emph{23}, 1778--1783\relax
\mciteBstWouldAddEndPuncttrue
\mciteSetBstMidEndSepPunct{\mcitedefaultmidpunct}
{\mcitedefaultendpunct}{\mcitedefaultseppunct}\relax
\EndOfBibitem
\bibitem[Yang et~al.(2007)Yang, Dammer, Bremond, Zandvliet, Kooij, and
  Lohse]{yang2007}
Yang,~S.; Dammer,~S.~M.; Bremond,~N.; Zandvliet,~H. J.~W.; Kooij,~E.~S.;
  Lohse,~D. \emph{Langmuir} \textbf{2007}, \emph{23}, 7072--7077\relax
\mciteBstWouldAddEndPuncttrue
\mciteSetBstMidEndSepPunct{\mcitedefaultmidpunct}
{\mcitedefaultendpunct}{\mcitedefaultseppunct}\relax
\EndOfBibitem
\bibitem[Seddon et~al.(2011)Seddon, Kooij, Poelsema, Zandvliet, and
  Lohse]{seddon2010c}
Seddon,~J. R.~T.; Kooij,~E.~S.; Poelsema,~B.; Zandvliet,~H. J.~W.; Lohse,~D.
  \emph{Phys. Rev. Lett.} \textbf{2011}, \emph{106}, 056101\relax
\mciteBstWouldAddEndPuncttrue
\mciteSetBstMidEndSepPunct{\mcitedefaultmidpunct}
{\mcitedefaultendpunct}{\mcitedefaultseppunct}\relax
\EndOfBibitem
\bibitem[Hampton and Nguyen(2010)Hampton, and Nguyen]{hampton2010}
Hampton,~M.~A.; Nguyen,~A.~V. \emph{Adv. Coll. Int. Sci.} \textbf{2010},
  \emph{154}, 30--55\relax
\mciteBstWouldAddEndPuncttrue
\mciteSetBstMidEndSepPunct{\mcitedefaultmidpunct}
{\mcitedefaultendpunct}{\mcitedefaultseppunct}\relax
\EndOfBibitem
\bibitem[com()]{comment_liquid}
Although the authors of Ref. \cite{simonsen2004} present evidence for
  nanobubbles in pure alcohols, this has never been reproduced.\relax
\mciteBstWouldAddEndPunctfalse
\mciteSetBstMidEndSepPunct{\mcitedefaultmidpunct}
{}{\mcitedefaultseppunct}\relax
\EndOfBibitem
\bibitem[Agrawal et~al.(2005)Agrawal, Park, Ryu, Hammond, Russell, and
  McKinley]{agrawal2005}
Agrawal,~A.; Park,~J.; Ryu,~D.~Y.; Hammond,~P.~T.; Russell,~T.~P.;
  McKinley,~G.~H. \emph{Nano Lett.} \textbf{2005}, \emph{5}, 1751--1756\relax
\mciteBstWouldAddEndPuncttrue
\mciteSetBstMidEndSepPunct{\mcitedefaultmidpunct}
{\mcitedefaultendpunct}{\mcitedefaultseppunct}\relax
\EndOfBibitem
\bibitem[Kameda and Nakabayashi(2008)Kameda, and Nakabayashi]{kameda2008}
Kameda,~N.; Nakabayashi,~S. \emph{Chem. Phys. Lett.} \textbf{2008}, \emph{461},
  122--126\relax
\mciteBstWouldAddEndPuncttrue
\mciteSetBstMidEndSepPunct{\mcitedefaultmidpunct}
{\mcitedefaultendpunct}{\mcitedefaultseppunct}\relax
\EndOfBibitem
\bibitem[Borkent et~al.(2009)Borkent, Sch\"onherr, Ca\"er, Dollet, and
  Lohse]{borkent2009}
Borkent,~B.~M.; Sch\"onherr,~H.; Ca\"er,~G.~L.; Dollet,~B.; Lohse,~D.
  \emph{Phys. Rev. E} \textbf{2009}, \emph{80}, 036315\relax
\mciteBstWouldAddEndPuncttrue
\mciteSetBstMidEndSepPunct{\mcitedefaultmidpunct}
{\mcitedefaultendpunct}{\mcitedefaultseppunct}\relax
\EndOfBibitem
\bibitem[Lou et~al.(2002)Lou, Gao, Xiao, Li, Li, Zhang, Li, Sun, Li, and
  Hu]{lou2002}
Lou,~S.; Gao,~J.; Xiao,~X.; Li,~X.; Li,~G.; Zhang,~Y.; Li,~M.; Sun,~J.; Li,~X.;
  Hu,~J. \emph{Materials Characterization} \textbf{2002}, \emph{48},
  211--214\relax
\mciteBstWouldAddEndPuncttrue
\mciteSetBstMidEndSepPunct{\mcitedefaultmidpunct}
{\mcitedefaultendpunct}{\mcitedefaultseppunct}\relax
\EndOfBibitem
\bibitem[Switkes and Ruberti(2004)Switkes, and Ruberti]{switkes2004}
Switkes,~M.; Ruberti,~J.~W. \emph{App. Phys. Lett.} \textbf{2004}, \emph{84},
  4759--4761\relax
\mciteBstWouldAddEndPuncttrue
\mciteSetBstMidEndSepPunct{\mcitedefaultmidpunct}
{\mcitedefaultendpunct}{\mcitedefaultseppunct}\relax
\EndOfBibitem
\bibitem[Zhang et~al.(2006)Zhang, Li, Maeda, and Hu]{zhang2006}
Zhang,~X.~H.; Li,~G.; Maeda,~N.; Hu,~J. \emph{Langmuir} \textbf{2006},
  \emph{22}, 9238--9243\relax
\mciteBstWouldAddEndPuncttrue
\mciteSetBstMidEndSepPunct{\mcitedefaultmidpunct}
{\mcitedefaultendpunct}{\mcitedefaultseppunct}\relax
\EndOfBibitem
\bibitem[Kameda et~al.(2008)Kameda, Sogoshi, and Nakabayashi]{kameda2008a}
Kameda,~N.; Sogoshi,~N.; Nakabayashi,~S. \emph{Surf. Sci.} \textbf{2008},
  \emph{602}, 1579--1584\relax
\mciteBstWouldAddEndPuncttrue
\mciteSetBstMidEndSepPunct{\mcitedefaultmidpunct}
{\mcitedefaultendpunct}{\mcitedefaultseppunct}\relax
\EndOfBibitem
\bibitem[Yang et~al.(2003)Yang, Duan, Fornasiero, and Ralston]{yang2003}
Yang,~J.; Duan,~J.; Fornasiero,~D.; Ralston,~J. \emph{J. Phys. Chem. B}
  \textbf{2003}, \emph{107}, 6139--6147\relax
\mciteBstWouldAddEndPuncttrue
\mciteSetBstMidEndSepPunct{\mcitedefaultmidpunct}
{\mcitedefaultendpunct}{\mcitedefaultseppunct}\relax
\EndOfBibitem
\bibitem[Miller et~al.(1999)Miller, Hu, Veeramasuneni, and Lu]{miller1999}
Miller,~J.~D.; Hu,~Y.; Veeramasuneni,~S.; Lu,~Y. \emph{Colloids and Surfaces A}
  \textbf{1999}, \emph{154}, 137--147\relax
\mciteBstWouldAddEndPuncttrue
\mciteSetBstMidEndSepPunct{\mcitedefaultmidpunct}
{\mcitedefaultendpunct}{\mcitedefaultseppunct}\relax
\EndOfBibitem
\bibitem[Zhang et~al.(2006)Zhang, Zhang, Zhang, Li, Shen, Ye, Fan, Fang, and
  Hu]{zhang2006b}
Zhang,~L.; Zhang,~Y.; Zhang,~X.; Li,~Z.; Shen,~G.; Ye,~M.; Fan,~C.; Fang,~H.;
  Hu,~J. \emph{Langmuir} \textbf{2006}, \emph{22}, 8109--8113\relax
\mciteBstWouldAddEndPuncttrue
\mciteSetBstMidEndSepPunct{\mcitedefaultmidpunct}
{\mcitedefaultendpunct}{\mcitedefaultseppunct}\relax
\EndOfBibitem
\bibitem[Yang et~al.(2009)Yang, Tsai, Kooij, Prosperetti, Zandvliet, and
  Lohse]{yang2009}
Yang,~S.; Tsai,~P.; Kooij,~E.~S.; Prosperetti,~A.; Zandvliet,~H. J.~W.;
  Lohse,~D. \emph{Langmuir} \textbf{2009}, \emph{25}, 1466--1474\relax
\mciteBstWouldAddEndPuncttrue
\mciteSetBstMidEndSepPunct{\mcitedefaultmidpunct}
{\mcitedefaultendpunct}{\mcitedefaultseppunct}\relax
\EndOfBibitem
\bibitem[Zhang et~al.(2010)Zhang, Zhang, Zhang, Hu, and Fang]{zhang2010}
Zhang,~L.; Zhang,~X.; Zhang,~Y.; Hu,~J.; Fang,~H. \emph{Soft Matter}
  \textbf{2010}, \emph{6}, 4515--4519\relax
\mciteBstWouldAddEndPuncttrue
\mciteSetBstMidEndSepPunct{\mcitedefaultmidpunct}
{\mcitedefaultendpunct}{\mcitedefaultseppunct}\relax
\EndOfBibitem
\bibitem[in~Chief: David R.~Lide(2005)]{crc}
in~Chief: David R.~Lide,~E. \emph{Handbook of Chemistry and Physics}, 86th ed.;
  Taylor and Francis, 2005\relax
\mciteBstWouldAddEndPuncttrue
\mciteSetBstMidEndSepPunct{\mcitedefaultmidpunct}
{\mcitedefaultendpunct}{\mcitedefaultseppunct}\relax
\EndOfBibitem
\bibitem[Rathgen(2008)]{helmutthesis}
Rathgen,~H. Superhydrophobic surfaces: From fluid mechanics to optics. Ph.D.\
  thesis, University of Twente, 2008\relax
\mciteBstWouldAddEndPuncttrue
\mciteSetBstMidEndSepPunct{\mcitedefaultmidpunct}
{\mcitedefaultendpunct}{\mcitedefaultseppunct}\relax
\EndOfBibitem
\bibitem[Seddon et~al.(2010)Seddon, Bliznyuk, Kooij, Poelsema, Zandvliet, and
  Lohse]{seddon2010a}
Seddon,~J. R.~T.; Bliznyuk,~O.; Kooij,~E.~S.; Poelsema,~B.; Zandvliet,~H.
  J.~W.; Lohse,~D. \emph{Langmuir} \textbf{2010}, \emph{26}, 9640--9644\relax
\mciteBstWouldAddEndPuncttrue
\mciteSetBstMidEndSepPunct{\mcitedefaultmidpunct}
{\mcitedefaultendpunct}{\mcitedefaultseppunct}\relax
\EndOfBibitem
\bibitem[Zhang et~al.(2005)Zhang, Li, Wu, Zhang, and Hu]{zhang2005}
Zhang,~X.~H.; Li,~G.; Wu,~Z.~H.; Zhang,~X.~D.; Hu,~J. \emph{Chin. Phys.}
  \textbf{2005}, \emph{14}, 1774--1778\relax
\mciteBstWouldAddEndPuncttrue
\mciteSetBstMidEndSepPunct{\mcitedefaultmidpunct}
{\mcitedefaultendpunct}{\mcitedefaultseppunct}\relax
\EndOfBibitem
\bibitem[Dunne et~al.(1996)Dunne, Mariwala, Rao, Sircar, Gorte, and
  Myers]{dunne1996}
Dunne,~J.~A.; Mariwala,~R.; Rao,~M.; Sircar,~S.; Gorte,~R.~J.; Myers,~A.~L.
  \emph{Langmuir} \textbf{1996}, \emph{12}, 5888--5895\relax
\mciteBstWouldAddEndPuncttrue
\mciteSetBstMidEndSepPunct{\mcitedefaultmidpunct}
{\mcitedefaultendpunct}{\mcitedefaultseppunct}\relax
\EndOfBibitem
\bibitem[Staudt et~al.(2005)Staudt, Herbst, Beutekamp, and Harting]{staudt2005}
Staudt,~R.; Herbst,~A.; Beutekamp,~S.; Harting,~P. \emph{Adsorption}
  \textbf{2005}, \emph{11}, 379--384\relax
\mciteBstWouldAddEndPuncttrue
\mciteSetBstMidEndSepPunct{\mcitedefaultmidpunct}
{\mcitedefaultendpunct}{\mcitedefaultseppunct}\relax
\EndOfBibitem
\bibitem[Sudik et~al.(2005)Sudik, Millward, Ockwig, Cote, Kim, and
  Yaghi]{sudik2005}
Sudik,~A.~C.; Millward,~A.~R.; Ockwig,~N.~W.; Cote,~A.~P.; Kim,~J.;
  Yaghi,~O.~M. \emph{J. Am. Chem. Soc.} \textbf{2005}, \emph{127},
  7110--7118\relax
\mciteBstWouldAddEndPuncttrue
\mciteSetBstMidEndSepPunct{\mcitedefaultmidpunct}
{\mcitedefaultendpunct}{\mcitedefaultseppunct}\relax
\EndOfBibitem
\bibitem[Doshi et~al.(2005)Doshi, Watkins, Israelachvili, and
  Majewski]{doshi2005}
Doshi,~D.~A.; Watkins,~E.~B.; Israelachvili,~J.~N.; Majewski,~J. \emph{PNAS}
  \textbf{2005}, \emph{102}, 9458--9462\relax
\mciteBstWouldAddEndPuncttrue
\mciteSetBstMidEndSepPunct{\mcitedefaultmidpunct}
{\mcitedefaultendpunct}{\mcitedefaultseppunct}\relax
\EndOfBibitem
\bibitem[Dammer and Lohse(2006)Dammer, and Lohse]{dammer2006}
Dammer,~S.~M.; Lohse,~D. \emph{Phys. Rev. Lett.} \textbf{2006}, \emph{96},
  206101\relax
\mciteBstWouldAddEndPuncttrue
\mciteSetBstMidEndSepPunct{\mcitedefaultmidpunct}
{\mcitedefaultendpunct}{\mcitedefaultseppunct}\relax
\EndOfBibitem
\bibitem[Mezger et~al.(2006)Mezger, Reichert, Sch\"oder, Okasinski, Schr\"oder,
  Dosch, Palms, Ralston, and Honkim\"aki]{mezger2006}
Mezger,~M.; Reichert,~H.; Sch\"oder,~S.; Okasinski,~J.; Schr\"oder,~H.;
  Dosch,~H.; Palms,~D.; Ralston,~J.; Honkim\"aki,~V. \emph{Proc. Nat. Acad.
  Sci.} \textbf{2006}, \emph{103}, 18401--18404\relax
\mciteBstWouldAddEndPuncttrue
\mciteSetBstMidEndSepPunct{\mcitedefaultmidpunct}
{\mcitedefaultendpunct}{\mcitedefaultseppunct}\relax
\EndOfBibitem
\bibitem[Sendner et~al.(2009)Sendner, Horinek, Bocquet, and Netz]{sendner2009}
Sendner,~C.; Horinek,~D.; Bocquet,~L.; Netz,~R.~R. \emph{Langmuir}
  \textbf{2009}, \emph{25}, 10768--10781\relax
\mciteBstWouldAddEndPuncttrue
\mciteSetBstMidEndSepPunct{\mcitedefaultmidpunct}
{\mcitedefaultendpunct}{\mcitedefaultseppunct}\relax
\EndOfBibitem
\bibitem[Zhang et~al.(2007)Zhang, Khan, and Ducker]{zhang2007}
Zhang,~X.~H.; Khan,~A.; Ducker,~W.~A. \emph{Phys. Rev. Lett.} \textbf{2007},
  \emph{98}, 136101\relax
\mciteBstWouldAddEndPuncttrue
\mciteSetBstMidEndSepPunct{\mcitedefaultmidpunct}
{\mcitedefaultendpunct}{\mcitedefaultseppunct}\relax
\EndOfBibitem
\bibitem[Weijs et~al.(2011)Weijs, Marchand, Andreotti, Lohse, and
  Snoeijer]{weijs2010}
Weijs,~J.~H.; Marchand,~A.; Andreotti,~B.; Lohse,~D.; Snoeijer,~J.~H.
  \emph{Phys. Fluids} \textbf{2011}, \emph{23}, 022001\relax
\mciteBstWouldAddEndPuncttrue
\mciteSetBstMidEndSepPunct{\mcitedefaultmidpunct}
{\mcitedefaultendpunct}{\mcitedefaultseppunct}\relax
\EndOfBibitem
\bibitem[Checco et~al.(2003)Checco, Guenoun, and Daillant]{checco2003}
Checco,~A.; Guenoun,~P.; Daillant,~J. \emph{Phys. Rev. Lett.} \textbf{2003},
  \emph{91}, 186101\relax
\mciteBstWouldAddEndPuncttrue
\mciteSetBstMidEndSepPunct{\mcitedefaultmidpunct}
{\mcitedefaultendpunct}{\mcitedefaultseppunct}\relax
\EndOfBibitem
\end{mcitethebibliography}

\providecommand*\mcitethebibliography{\thebibliography}
\csname @ifundefined\endcsname{endmcitethebibliography}
  {\let\endmcitethebibliography\endthebibliography}{}

\end{document}